\documentclass[fleqn,usenatbib,useAMS]{mnras}

\DeclareRobustCommand{\VAN}[3]{#2}
\let\VANthebibliography\thebibliography
\def\thebibliography{\DeclareRobustCommand{\VAN}[3]{##3}\VANthebibliography}

\usepackage{xcolor}
\usepackage{ulem}
\usepackage{graphicx}	% Including figure files
\usepackage{subfigure}	
\usepackage{amsmath}	% Advanced maths commands
\usepackage{multicol}   % Multi-column entries in tables
\usepackage{bm}	        % Bold maths symbols, including upright Greek
\usepackage{pdflscape}	% Landscape pages
\usepackage{comment}
\usepackage{bmpsize}
\usepackage{grffile}

\newcommand{\Mh}{M_{\rm h}}
\newcommand\Msun{M_{\odot}}
\newcommand\Mmin{M_{\min}}
\newcommand\Mstar{M_{\star}}

\newcommand\fsat{f_{\rm s}}
\newcommand\bg{b_{\rm g}}

\newcommand\sigmalogM{\sigma_{\log{M}}}

\newcommand{\Nc}{N_{\rm c}}
\newcommand{\Ns}{N_{\rm s}}

\usepackage[T1]{fontenc}
\usepackage{ae,aecompl}

\usepackage{newtxtext,newtxmath}

\title[HSC lightcone mock catalogues for cosmological analysis]{{\sc Mock Observatory}: two thousand lightcone mock catalogues of luminous red galaxies from the Hyper Suprime-Cam Survey for the cosmological large-scale analysis}

\author[S. Ishikawa, T. Okumura, and T. Nishimichi]{
Shogo Ishikawa,$^{1,2,3}$\footnotemark{\thanks{shogo.ishikawa.astro@gmail.com, shogo.ishikawa@yukawa.kyoto-u.ac.jp}}
Teppei Okumura$^{4,5}$ 
and
Takahiro Nishimichi$^{6,1,5}$
\\
$^{1}$Center for Gravitational Physics and Quantum Information, Yukawa Institute for Theoretical Physics, Kyoto University, Sakyo-ku, Kyoto 606-8502, Japan\\
$^{2}$Center for Computational Astrophysics, National Astronomical Observatory of Japan, Mitaka, Tokyo 181-8588, Japan\\
$^{3}$National Astronomical Observatory of Japan, Mitaka, Tokyo 181-8588, Japan\\
$^{4}$Institute of Astronomy and Astrophysics, Academia Sinica, No.~1, Section~4, Roosevelt Road, Taipei 10617, Taiwan\\
$^{5}$Kavli Institute for the Physics and Mathematics of the Universe (WPI), UTIAS, The University of Tokyo, Kashiwa, Chiba 277-8583, Japan\\
$^{6}$Department of Astrophysics and Atmospheric Sciences, Faculty of Science, Kyoto Sangyo University, Motoyama, Kamigamo, Kita-ku, Kyoto 603-8555, Japan
}

\date{Accepted XXX. Received YYY; in original form ZZZ}

\pubyear{2023}

\begin{document}
\label{firstpage}
\pagerange{\pageref{firstpage}--\pageref{lastpage}}
\maketitle

\begin{abstract}
Estimating a reliable covariance matrix for correlation functions of galaxies is a crucial task to obtain accurate cosmological constraints from galaxy surveys.
We generate $2,000$ independent lightcone mock luminous red galaxy (LRGs) catalogues at $0.3 \leq z \leq 1.25$, designed to cover CAMIRA LRGs observed by the Subaru Hyper Suprime-Cam Subaru Strategic Programme (HSC SSP). 
We first produce full-sky lightcone halo catalogues using a COmoving Lagrangian Acceleration (COLA) technique, and then trim them to match the footprints of the HSC SSP S20A Wide layers. 
The mock LRGs are subsequently populated onto the trimmed halo catalogues according to the halo occupation distribution model constrained by the observed CAMIRA LRGs. 
The stellar mass ($\Mstar$) is assigned to each LRG by the subhalo abundance-matching technique using the observed stellar-mass functions of CAMIRA LRGs. 
We evaluate photometric redshifts (photo-$z$) of mock LRGs by incorporating the photo-$z$ scatter, which is derived from the observed $\Mstar$--photo-$z$-scatter relations of the CAMIRA LRGs. 
We validate the constructed full-sky halo and lightcone LRG mock catalogues by comparing their angular clustering statistics (i.e., power spectra and correlation functions) with those measured from the halo catalogues of full $N$-body simulations and the CAMIRA LRG catalogues from the HSC SSP, respectively. 
We detect clear signatures of baryon acoustic oscillations (BAOs) from our mock LRGs, whose angular scales are well consistent with theoretical predictions. 
These results demonstrate that our mock LRGs can be used to evaluate covariance matrices at large scales and provide predictions for the BAO detectability and cosmological constraints. 
\end{abstract} 


\section{Introduction} \label{sec:intro}
In recent decades, our understanding of the large-scale structure of the Universe has dramatically advanced in both theoretical and observational aspects. 
On the theoretical side, significant progress has been made through massively parallel $N$-body simulations of dark matter clustering, providing insights into structure formation across cosmic time over a wide dynamic range \citep[e.g.,][]{springel05,ishiyama15,ishiyama21}. 
Furthermore, the availability of large computational resources has facilitated studies of baryonic effects and their evolution within cosmological volumes through hydrodynamical simulations \citep[e.g.,][]{genel14,schaye15,springel18}. 
Besides the theoretical achievements, precise 3D galaxy maps have been drawn by extensive redshift surveys, offering valuable opportunities to investigate structure formation scenarios and the nature of dark matter and dark energy \citep[e.g.,][]{york00,lefevre05,drinkwater10,newman13,tonegawa15}. 
Analysing these data at cosmological scales has provided insight into the expansion and growth histories of the Universe by detecting signals of baryon acoustic oscillations (BAOs) and redshift-space distortions \citep[e.g.,][]{hawkins03,eisenstein05,tegmark06,guzzo08,okumura08,percival10,reid10,beutler11,okumura16}.  

In addition to spectroscopic redshift surveys mentioned above, recent ground-based large telescopes also have provided multi-wavelength photometric data covering huge survey volumes. 
Notable examples include the Hyper Suprime-Cam Subaru Strategic Programme \citep[HSC SSP;][]{aihara18}, the Dark Energy Survey \citep[DES;][]{des16}, and the Kilo-Degree Survey \citep[KiDS;][]{kids13}. 
However, extracting cosmological information from extensive galaxy surveys poses certain challenges. 
One of the primary difficulties in cosmological analyses arises in the estimation of covariance matrices. 
They are essential ingredients in the extraction of cosmological parameters from observational data, and thus an accurate covariance estimation is needed to ensure the reliability and robustness of the results \citep[e.g.,][]{percival14}.

The jackknife (JK) resampling technique is one of the simplest methods used to evaluate covariance matrices \citep{shao89,norberg09}. 
It is widely employed in studies of galaxy clustering at small to intermediate scales \citep[e.g.,][]{zehavi11,ishikawa20,okumura21}. 
However, \citet{shirasaki17} revealed that the covariance matrices estimated by the JK method tend to be underestimated at scales larger than the sub-regions removed in the JK resampling process. 
%Some of the observational studies have used the JK resampling method to assess the covariance matrices at the BAO scale, but 
Thus, the JK method can be employed to evaluate covariance matrices only on scales smaller than the JK sub-regions. 
This makes it challenging to estimate the covariance matrices for the analysis of large-scale structure such as BAO. 

Constructing many independent mock galaxy catalogues that accurately mimic the observed galaxy distributions offers an alternative approach to estimate covariance matrices. 
The covariance matrix from such independent catalogues can provide a reliable estimator of the cosmic variance, unlike the JK method, which relies on a observed single data set. 

The best way to construct multiple mock galaxy catalogues is to perform cosmological $N$-body simulations with independent initial conditions and subsequently assign galaxies and their baryonic information to the dark matter haloes in post processes \citep[e.g.,][]{merson13,crocce15}. 
However, it is generally necessary to create several thousand mock galaxy catalogues for the estimation of the covariance matrix, which inevitably incurs substantial computational costs to perform full $N$-body simulations with multiple realisations, large simulation volumes, and high mass resolutions for comparison with survey data. 
To make the problem tractable, various approximate computational techniques have been developed to expedite the generation of independent mock catalogues. 

The second-order perturbation theory \citep[2LPT;][]{bouchet95,scoccimarro98,bernardeau02,crocce06} is one of the simplest techniques used to construct multiple halo catalogues. 
The 2LPT computes the displacement field from the initial Lagrangian positions to the final Eulerian positions that takes into account up to the second order terms in the perturbative expansion \citep[cf.][]{jenkins10}. 
Using the 2LPT framework to generate a matter density field, \citet{manera13} constructed $600$ mock catalogues of CMASS galaxy samples by calibrating halo mass functions from $N$-body simulations. 
Similarly, \citet{avila18} employed the HALOGEN method, which is based upon the 2LPT method, to generate $1,800$ mock galaxy catalogues for the DES Year-1 BAO samples. 
Although the 2LPT technique allows fast generation of matter distributions, it is important to note that the accuracy of the resulting matter distribution is degraded compared to full $N$-body simulation, even in the weakly nonlinear regime~\citep[see][]{tassev13,koda16}. 

In order to achieve fast and accurate computations at a reduced computational cost, \citet{tassev13} developed a COmoving Lagrangian Acceleration (COLA) technique. 
This method allows us to generate relatively accurate matter distributions using a smaller number of time steps compared to full $N$-body simulations. 
The COLA technique refines the matter distributions generated by the 2LPT method by computing the residual gravitational force using the Particle Mesh (PM) solver and capable of evolving matter distribution at linear to quasi-linear scales. 
This approach has been successfully utilised in recent large galaxy surveys to construct multiple independent mock galaxy catalogues for large-scale analyses. 
For instance, \citet{koda16} developed $600$ mock galaxy catalogues for the WiggleZ Dark Energy Survey \footnote{https://github.com/junkoda/cola\_halo}. 
Additionally, \citet{izard16} and \citet{ferrero21} generated $488$ mock galaxy catalogues for the DES Year-3, further demonstrating the advantages of the COLA method in efficiently generating mock galaxy catalogues with sufficient accuracy for performing BAO scale analyses. 
Apart from the studies on BAO scales, the GLAM $N$-body approach \citep{klypin18} was used to evaluate covariance matrices of DESI-like LRGs \citep{hernandez21}. 

In this paper, we present $2,000$ lightcone mock catalogues of luminous red galaxies (LRGs) selected from the HSC SSP. 
The main objective of this study is to calculate covariance matrices for large-scale cosmological analyses. 
To generate these mock LRG catalogues, we employ the COLA method and create $2,000$ independent realisations of mock halo catalogues. 
LRGs are then assigned to each halo catalogue using a hybrid technique that combines the halo occupation distribution \citep[HOD; e.g.,][]{berlind02,kravtsov04} and the subhalo abundance-matching method \citep[SHAM; e.g.,][]{vale04,conroy06}. 

The full-sky mock halo catalogues with independent $108$ realisations presented by \citet{takahashi17} are widely used to compare with the observational data from the HSC SSP, particularly in weak-lensing studies \citep[e.g.,][]{hamana20,shirasaki21,miyatake22}. 
However, the minimum halo masses in \citet{takahashi17} exceed the observed $\Mmin$ parameters of the LRGs selected from the HSC SSP Wide layer at $z \geq 0.55$ \citep{ishikawa21}. 
As a result, it is not possible to construct realistic mock LRG catalogues that cover the full ranges of observed dynamics. 
Our mock LRG catalogues address this issue by satisfying the observed minimum halo masses within the redshift range of at $0.30 \leq z \leq 1.25$ and matching the footprints of the HSC SSP Wide layers. 
This allows a direct comparison with observations and enables the prediction of the statistical significance of the BAO detection. 
To achieve the construction of realistic mock LRG catalogues, we incorporate observational constraints on HOD-model parameters, stellar mass functions, and uncertainties on photometric redshifts. 

This paper is organized as follows. 
In Section~2, we show the properties of our dark-matter-only simulations using the COLA method and give details of the observational results that are used to generate mock LRG catalogues in Section~3. 
The methodology to construct lightcone mock LRG catalogues from discrete snapshots is given in Section~4. 
We check the validity of our mocks by calculating the two-point statistical quantities and the redshift distributions, and compare them with those from the full $N$-body simulations by \citet{takahashi17} and the observations by \citet{ishikawa21} in Section~5. 
In Section~6, we give summary and conclusion of this paper. 

Throughout this paper, we employ the cosmological parameters derived from the observation of the cosmic microwave background by the Planck satellite \citep{planck15}; that is, the cosmic density parameters are $\Omega_{\rm m} = 0.3089$, $\Omega_{\rm \Lambda} = 0.6911$, and $\Omega_{\rm b} = 0.049$, the dimensionless Hubble parameter is $h=0.6774$, the matter fluctuation averaged over $8$ $h^{-1}$Mpc is $\sigma_{8} = 0.8159$, and the spectral index is $n_{\rm s} = 0.9667$, respectively. 
Stellar and dark halo masses are denoted by $\Mstar$ and $\Mh$, and are scaled in units of $h^{-2}\Msun$ and $h^{-1}\Msun$, respectively. 
All the logarithms in this paper are base $10$. 

\section{COLA Simulations} \label{sec:simulation}
Cosmological analysis of galaxy clustering typically requires thousands of mock galaxy catalogues to estimate a reliable covariance matrix \citep[e.g.,][]{dodelson13,kitaura16}. 
We aim to generate $2,000$ independent realisations of mock galaxy distributions. 
\citet{takahashi09} estimated dispersions among power spectrum covariance matrices using a subset of the full $5,000$ realisations. 
The analysis revealed that the dispersions of the diagonal elements followed a scaling relation of $2/N_{r}$, where $N_{r}$ represents the number of realisations, and reached an accuracy of $\sim0.1\%$ with $N_r=1,000$ for the diagonal elements, which would be sufficient for the actual correlation-function analyses from existing or near-future surveys. 
% Assuming a similar behaviour for correlation functions, increasing the number of realisations to $5,000$ would only marginally improve the dispersion by several percentage points; hence, we conclude 
Following this result, we here employ $2,000$ realisations to be conservative.
% are sufficient to obtain covariances 
% with moderate precision.
To achieve this with reasonable computational cost, we run dark-matter-only simulations with the COLA method that is based upon  the 2LPT \citep[][see also \citealp{koda16} for more detail]{tassev13}, instead of carrying out full $N$-body simulations, which calculate the gravitational force on the simulation particles from the mass distribution typically more than 1,000 times with small time intervals. 

\subsection{Dark matter distributions}\label{subsec:nbody_sim}

We use a publicly available simulation code, {\sc L-PICOLA}\footnote{https://cullanhowlett.github.io/l-picola/} \citep{howlett15}, which is an extended version of the original code. 
The {\sc L-PICOLA} is designed to enable massively parallel computations using Message Passing Interface (MPI) and has the ability to generate an on-the-fly lightcone output from simulation particles. 
However, we do not use the latter option in our simulations and store only constant-time snapshots of mass distribution in the {\sc Gadget-2} format to reduce date storage consumption. 
We create lightcones only for halos and galaxies as a postprocess (see Section~\ref{sec:lightcone}). 
Refer to \citet{howlett15} for a comparison of the accuracy and speed of {\sc L-PICOLA} with full $N$-body simulations. 

We perform the COLA simulations on the Aterui II supercomputer (Cray XC50) operated by Center for Computational Astrophysics (CfCA), National Astronomical Observatory of Japan (NAOJ). 
We employ $N_{\rm p} = 2,560^{3}$ dark matter particles in a periodic comoving box with the side length of $L_{\rm box}=2,048$ $h^{-1}$Mpc and the PM force resolution is set by $N_{\rm mesh} = 2,560$ mesh cells along each side of the simulation box. 

We choose the initial redshift to be $z_{\rm init} = 49.0$, and the number of time step from $z_{\rm init}$ to $z=0.0$ to be 100. 
The particle distributions are extracted at $z=1.4000$, $1.0000$, $0.7143$, $0.5000$, $0.3333$, $0.2000$, $0.0909$, and $0.0000$ that are determined by equal divisions of scale factors corresponding to redshifts from $z=1.4$ to $z=0.0$. 
It is noted that increasing the number of outputting snapshots from $8$ to $15$ does not significantly improve the accuracy of halo positions. 
We use $27$ computational nodes and $1,080$ CPU cores of the Cray XC50 system, and each COLA simulation takes about $3$ hours. 

The initial conditions of the particle distributions are generated using a parallel code, {\sc 2LPTic}\footnote{https://cosmo.nyu.edu/roman/2LPT/} \citep{scoccimarro98,scoccimarro12}, run in an internal module of {\sc L-PICOLA}. 
The linear matter power spectrum at $z_{\rm init}$ is computed using the publicly available cosmological Boltzmann code, {\sc CAMB}\footnote{https://camb.readthedocs.io/} \citep{lewis00}. 
We adopt the linear time-step spacing in the scale factor with the modified COLA time-stepping parameter ${\rm nLPT}=-2.5$ \citep[see][]{howlett15}. 
Note that we consider only Gaussian initial conditions in this paper, whilst the code supports primordial non-Gaussianities. 

\subsection{Halo catalogues of discrete snapshots} \label{subsec:halocat}
\subsubsection{Halo identification and halo mass functions} \label{subsubsec:hmf}
\begin{figure}
\includegraphics[width=\columnwidth]{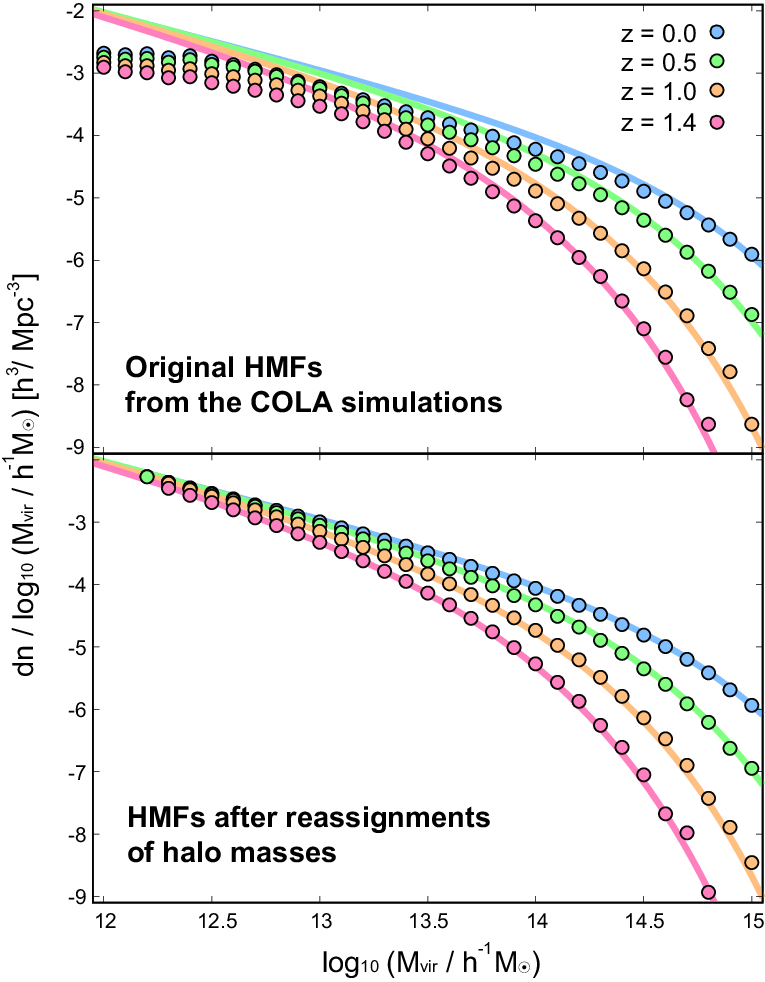}
\caption{Halo mass functions of one of the realisations constructed from COLA technique. The halo mass functions are calculated from snapshots at redshift $0.0$ (blue), $0.5$ (green), $1.0$ (orange), and $1.4$ (red). Solid lines represent analytical halo mass functions \citep{behroozi13}. The upper panel shows the original halo mass functions derived from the COLA simulations, whereas the lower panel are the halo mass functions after the halo-mass reassignments presented in Section~\ref{subsubsec:hmf}. Note that the halo mass functions of both panels are obtained from the same realisation. }
\label{fig:hmf}
\end{figure}

The dark halo catalogue at each snapshot is constructed using the publicly available phase-space temporal halo finder, {\sc ROCKSTAR}\footnote{https://bitbucket.org/gfcstanford/rockstar/} \citep{behroozi13_rockstar}. 
We employ a softening length of $\epsilon = 0.10$ $h^{-1}$Mpc, which is used as a minimum distance from halo centres in the calculation process of maximum circular velocities, and a friends-of-friends linking length of $b=0.2$ in the {\sc ROCKSTAR} algorithm. 
%{\bf We employ a friends-of-friends linking length of $b=0.2$ in the {\sc ROCKSTAR} algorithm. 
%We use the virial mass \citep{bryan98} listed in the catalogue for the subsequent analyses. 
The virial mass \citep{bryan98} listed in the catalogue is used for the subsequent analyses. 

The halo mass functions measured from the COLA simulations are systematically small ($\sim 10 \%$ suppression compared to the results of full $N$-body simulations with $\sim$ 0.3 dex scatter) even for relatively massive haloes with $\Mh > 10^{13}h^{-1}\Msun$ due to the lack of force and time resolution \citep[see][]{koda16}. 
The upper panel of Figure~\ref{fig:hmf} shows the halo mass function extracted from a random realisation of our COLA simulations at four different redshifts. 
Although the data points from the simulation are consistent with a fitting formula by \citet{behroozi13} at the massive end ($\Mh \lesssim 10^{14-14.5} h^{-1} \Msun$), the former is systematically suppressed towards lower masses, especially in the low-$z$ snapshots. 

To compensate for the deficiency, we reassign the masses of haloes generated by the COLA simulations to match the prediction of \citet{behroozi13}, which is an improved model of \citet{tinker08}. 
To do this, we first calculate the \citet{behroozi13} model at the redshift of each snapshot using the {\sc HMF}\footnote{https://hmf.readthedocs.io/} package \citep{murray13}. 
We then use the expected \textit{cumulative} number of halos in the simulated volume to create a mapping between the rank of the original halo mass and a new mass, reproducing the \citet{behroozi13} model. 
This procedure is similar to the abundance-matching technique that connects halo masses (or other proxies such as the maximum circular velocity) to galaxy stellar masses; however, our mass-matching method does not incorporate any scatter between halo masses from the COLA and the analytical formula. 
The mass-corrected halo mass functions are by construction in agreement with the analytical model down to the mass resolution limit, as shown in the lower panel of Figure~\ref{fig:hmf}. 

\subsubsection{Subhalo identification and ancestor--descendant relations} \label{subsubsec:subhalo}
Subhaloes are identified using the {\tt FindParents} algorithm in the {\sc ROCKSTAR} halo finder, and haloes are linked to their descendants/ancestors by an internal module of the {\sc ROCKSTAR} instead of constructing precise halo merger trees. 
The continuity of halo masses between adjacent snapshots is no longer conserved by the above mass-matching technique. 
Nevertheless, the merger tree is useful for connecting haloes between snapshots when they cross the observer's past lightcone. 
We find their positions and velocities on the lightcone by interpolating between the two snapshots that sandwich the lightcone crossing (see Section~\ref{subsec:halo_lightcone}). 

\section{Observation} \label{sec:observation}

\citet{ishikawa21} constrained HODs of the HSC LRGs based upon the angular correlation functions (ACFs). 
In this paper, we construct mock LRG catalogues that mimic their spatial distributions by populating the simulated haloes with galaxies according to the constrained HOD model. 
Section~\ref{subsec:obs_overview} briefly summarises the observational results of \citet{ishikawa21}. 

\subsection{Overview of observational data} \label{subsec:obs_overview}

The LRG samples were selected from the HSC SSP S16A Wide layer data covering over $\sim 124$ deg$^{2}$ \citep{aihara18} by the CAMIRA algorithm \citep{oguri14,oguri18}. 
Since the CAMIRA algorithm provides stellar mass and photometric redshift for each LRG, \citet{ishikawa21} constructed LRG subsamples at different redshifts and stellar masses. 
The LRGs were selected based on their redshift and optical colours \citep[see][for more details]{oguri18}, and further classification was made according to their redshifts and stellar masses for clustering analyses \citep{ishikawa21}:  
haloes with $0.1 \leq z \leq 1.05$ and $M_{\star} \geq 10^{10.0}h^{-2}\Msun$ were  divided into subsamples to investigate the dependence of the HOD parameters upon these properties. 

We adopted the standard HOD model proposed by \citet{zheng05} to study the HOD of our LRG subsamples. 
In this model, the expected total number of galaxies within a halo, $N_{\rm tot}$, is described by the sum of the number of central ($\Nc$) and satellite ($\Ns$) galaxies as a function of the halo mass, $\Mh$, 
\begin{equation}
N_{\rm tot}(\Mh) = \Nc(\Mh) \left[ 1 + \Ns(\Mh) \right],
\label{eq:Nt}
\end{equation}
where $\Nc$ and $\Ns$ are respectively given by 
\begin{align}
\Nc(\Mh) &= \frac{1}{2} \left[ 1 + {\rm erf}\left( \frac{\log_{10}{\Mh} - \log_{10}{\Mmin}}{\sigmalogM} \right) \right] ~,
\label{eq:Nc} \\
\Ns(\Mh) &= \left( \frac{\Mh - M_{0}}{M_{1}} \right)^{\alpha} ~.
\label{eq:Ns}
\end{align}
The actual number of central and satellite galaxies in a halo is assumed to follow a Bernoulli and a Poisson distribution, respectively, with the mean values as described in the equations above \citep[e.g.,][]{hikage13a,delgado22}. 
This HOD model contains five free parameters, $\Mmin$, $M_{1}$, $M_{0}$, $\sigmalogM$, and $\alpha$. 
They were constrained by fitting the observed ACFs and number density under flat priors. 
This model reproduces the observed ACFs of LRGs successfully, and helps us to infer the relationship between LRGs and host dark haloes, and its dependence on redshifts and stellar masses. 
See Figure 3 and Table 2 of \citet{ishikawa21} for the ACFs and the derived HOD parameter constraints, respectively. 

\subsection{Additional HOD analysis for higher-$z$ subsample} \label{subsec:obs_hod_high-z}

\begin{figure}
\includegraphics[width=\columnwidth]{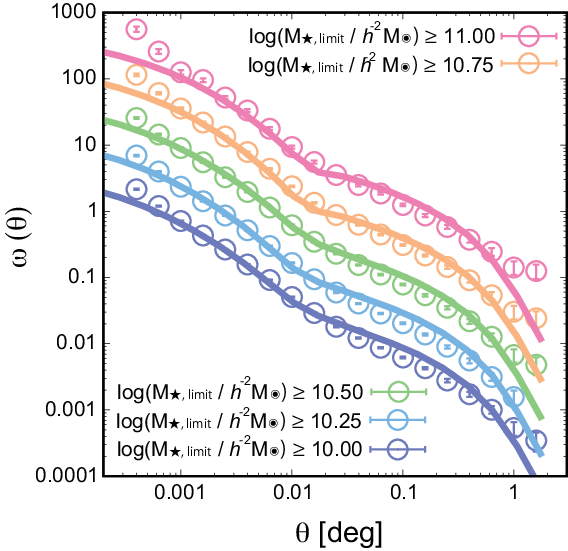}
\caption{Observed ACFs of CAMIRA LRGs (circles) and their best-fitting HOD model predictions (solid lines) at $1.05 \leq z \leq 1.25$. The HOD analysis is performed by dividing the LRGs into five cumulative stellar mass subsamples with $\Delta \Mstar = 0.25$ dex. Amplitudes of ACFs are shifted arbitrarily for an illustrative purpose.}
\label{fig:hod_z5}
\end{figure}

The upper bound of the photometric redshift of the LRGs constructed in \citet{ishikawa21} based on the HSC SSP S16A Wide layer was $z=1.05$. 
Since it extends to $z=1.25$ in the updated CAMIRA LRG catalogue from the S20A Wide layer \citep{aihara22}, we additionally make a higher-$z$ sample covering at $1.05 \leq z \leq 1.25$ and repeat the analysis described above. 

\begin{table*}
\caption{Best-fitting HOD Parameters of Cumulative Stellar-mass Limited CAMIRA LRGs at $1.05 \leq z \leq 1.25$}
\label{tab:hod_params}
\scalebox{0.92}{
\begin{tabular}{lcccccccccc}
\hline
$\log_{10}(M_{\star {\rm , limit}}^{a})$ & $\log_{10}(M_{\star {\rm , med}})$ & $n_{g}^{b}$ & $\log_{10}\Mmin$ & $\log_{10}M_{1}$ & $\log_{10}M_{0}$ & $\sigmalogM$ & $\alpha$ & $\fsat^{c}$ & $\bg^{d}$ & $\chi^{2}$/d.o.f. \\
\hline
$10.0$ & $10.2$ & $2.396$ & $12.566^{+0.038}_{-0.034}$ & $13.346^{+0.068}_{-0.060}$ & $8.049^{+2.051}_{-2.059}$ & $0.817^{+0.056}_{-0.169}$ & $1.044^{+0.071}_{-0.104}$ & $0.138 \pm 0.021$ & $1.827 \pm 0.024$ & $2.156$ \\
$10.25$ & $10.4$ & $2.139$ & $12.534^{+0.037}_{-0.069}$ & $13.382^{+0.075}_{-0.069}$ & $8.100^{+2.158}_{-2.088}$ & $0.726^{+0.107}_{-0.569}$ & $1.008^{+0.125}_{-0.129}$ & $0.142 \pm 0.050$ & $1.923 \pm 0.062$ & $1.567$ \\
$10.5$ & $10.5$  & $1.426$ & $12.700^{+0.036}_{-0.034}$ & $13.513^{+0.069}_{-0.067}$ & $8.636^{+2.455}_{-2.498}$ & $0.733^{+0.107}_{-0.356}$ & $1.099^{+0.106}_{-0.206}$ & $0.127 \pm 0.037$ & $2.014 \pm 0.053$ & $0.414$ \\
$10.75$ & $10.8$ & $0.746$ & $12.784^{+0.037}_{-0.032}$ & $13.992^{+0.108}_{-0.103}$ & $8.763^{+2.603}_{-2.598}$ & $0.557^{+0.214}_{-0.379}$ & $0.951^{+0.291}_{-0.272}$ & $0.087 \pm 0.034$ & $2.247 \pm 0.061$ & $0.171$ \\
$11.0$ & $11.1$ & $0.297$ & $13.079^{+0.039}_{-0.031}$ & $14.408^{+0.236}_{-0.172}$ & $9.172^{+2.833}_{-2.795}$ & $0.549^{+0.238}_{-0.436}$ & $1.036^{+0.303}_{-0.244}$ & $0.047 \pm 0.026$ & $2.598 \pm 0.088$ & $0.087$ \\
\hline
\multicolumn{9}{l}{\footnotesize All halo mass parameters are in units of $h^{-1}\Msun$. }\\
\multicolumn{9}{l}{\footnotesize$^a$ Threshold stellar mass of each subsample in units of $h^{-2}\Msun$ in a logarithmic scale.}\\
\multicolumn{9}{l}{\footnotesize$^b$ LRG number density in units of $10^{-3} h^{-3}$Mpc$^{3}$.}\\
\multicolumn{9}{l}{\footnotesize$^c$ Satellite fractions}\\ 
\multicolumn{9}{l}{\footnotesize$^d$ Galaxy biases}\\ 
\end{tabular}
}
\end{table*}

In Figure~\ref{fig:hod_z5}, we show the ACFs measured using the \citet{landy93} estimator for the stellar-mass subsamples at the additional higher-$z$ bin. 
As in \citet{ishikawa21}, the HOD modelling is performed using the CosmoPMC library \citep{Kilbinger11}. 
We assume the same analytical models as in \citet{ishikawa21}. Namely, we adopt the large-scale halo bias of \citet{tinker10}, the halo exclusion model of \citet{tinker05}, the halo radial profile of the NFW profile \citep{navarro97}, the non-linear power spectrum of \citet{smith03} with the transfer function of \citet{eisenstein98}, and the halo mass function of \citet{sheth01}. 
The photometric redshift distributions are assumed as selection functions of LRGs of each tomographic bin and the projected ACFs from the HOD model are computed by integrating along the selection functions. 
The best-fitting HOD model predictions are shown for each subsample by the solid curves in Figure~\ref{fig:hod_z5} and the constrained HOD parameters are summarised in Table~\ref{tab:hod_params}.
It is worth noting that the constraints on the HOD parameters for this redshift bin are obtained from a larger angular area of the HSC SSP S20A than those for $0.1\leq z \leq 1.05$ \citep{ishikawa21}. 
While the observed ACFs are generally well reproduced by the HOD model, there is a small excess in the model curves in the 2-halo regime, especially at $\theta \sim 0.1$. 
This trend was also found in previous studies of high-$z$ galaxies \citep[e.g.][]{ishikawa16,ishikawa17,harikane22}. 
Although the small disagreement could be resolved by incorporating the non-linear halo bias model to the analysis, we use the \citet{zheng05} model with the linear halo bias for consistency with our previous analysis of the lower-$z$ subsamples \citet{ishikawa21}. 

Finally, we have the best-fitting HOD parameters of the LRG samples for five tomographic bins, 
\begin{equation}
z \in [0.1, 0.3), [0.3, 0.55), [0.55, 0.8), [0.8, 1.05), [1.05, 1.25]. 
\label{eq:zbin}
\end{equation}
In the following sections, we reconstruct LRG distributions in the COLA simulations using the above HOD parameters. 

\section{Generating HSC LRG Lightcone Catalogue} \label{sec:lightcone}

We have developed a Python code suite, named {\sc Mock Observatory}, which generates lightcone mock galaxy catalogues covering an arbitrary survey footprint using halo catalogues from discrete snapshots. 
In this section, we present the procedures for creating full-sky lightcone halo catalogues, trimming them based upon the inputted survey geometries, and assigning galaxies to haloes using the HOD and SHAM techniques. 

\subsection{Full-sky lightcone dark halo catalogues} \label{subsec:halo_lightcone}

To generate lightcone mock LRG catalogues, we first construct full-sky lightcone dark halo catalogues from the {\sc L-PICOLA} halo catalogues discretely sampled at different redshifts identified by the {\sc ROCKSTAR} halo finder in Section~\ref{subsec:halocat}. 
The procedure of extracting the full-sky lightcone dark halo catalogues is as follows. 
For two subsequent snapshots at $z_i$ and $z_{i+1}$ ($0 \leq z_{i} < z_{i+1} \leq 1.4$), we (i) place replicated periodic boxes to cover the comoving volume up to $z=1.4$, (ii) identify dark haloes that cross the observer's past lightcone during the epoch between $z_{i}$ and $z_{i+1}$, (iii) place identified haloes at observer's past lightcone by interpolating the positions and velocities between snapshots, and (iv) repeat these processes until observer's lightcone is covered from $z=1.4$ to $0.0$. 

\subsubsection{Identifying haloes that cross observer's past lightcone} \label{subsubsec:identify_halo}
We identify haloes that have crossed the past lightcone of a hypothetical observer randomly located in the simulation box. 
We consider the Friedmann-Lema\^itre-Robertson-Walker (FLRW) metric:
\begin{equation}
ds^{2} = -c^{2}dt^{2} + a(t)^{2}[dr^{2} + r^{2}d\Omega^{2}], 
\label{eq:flrw_metric}
\end{equation}
where $a(t)=(1+z)^{-1}$ is the scale factor, and $dr$ and $d\Omega$ are the comoving distance and solid angle, respectively. 
For notational convenience, we take the observer's position to be at the origin in the following descriptions, although it is randomly chosen for each realisation in the actual mock making process. 
Then, those haloes that cross the observer's past lightcone between $z_i < z < z_{i+1}$ can be found by checking the sign of the line element, $ds^2$: they are time-like ($ds^{2}<0$) haloes at $z_{i+1}$ but space-like ($ds^{2}>0$) haloes at $z_{i}$. 

\begin{figure}
\includegraphics[width=\columnwidth]{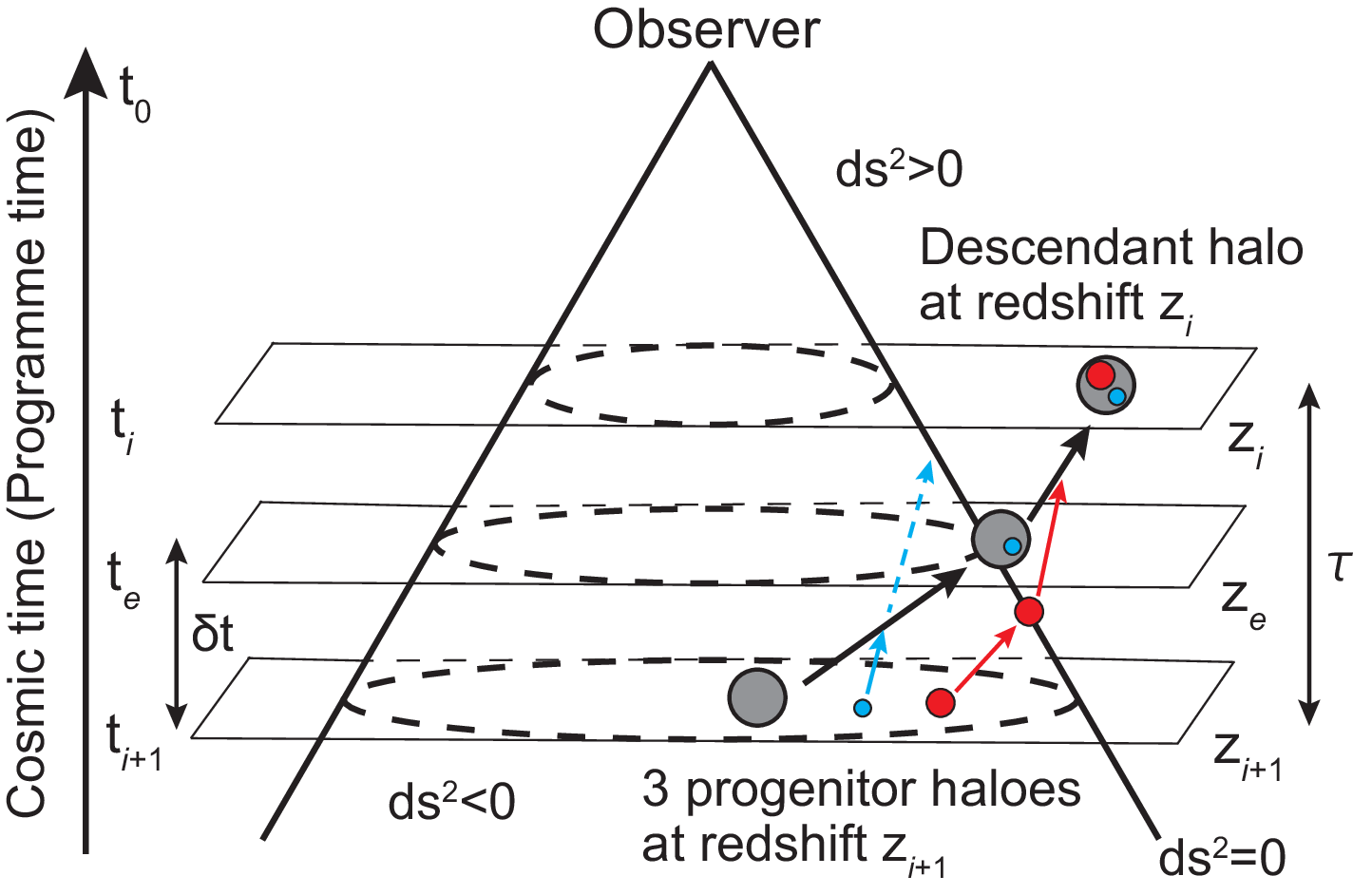}
\caption{Schematic view of haloes crossing the observer’s lightcone at time $t_{e}$. Time-like haloes at a snapshot $z_{i+1}$ cross the observer's past lightcone and become a space-like halo at an adjacent snapshot $z_{i}$. If a part of progenitor haloes merges before crossing the lightcone, they do not appear in the lightcone mock catalogue (small blue circle). On the other hand, progenitors that cross the lightcone before merging are listed in the lightcone mock as a separate halo (medium red circle). }
\label{fig:halo_crossing_time}
\end{figure}

Figure~\ref{fig:halo_crossing_time} shows an example of haloes crossing the observer's lightcone at time $t_{e}$. 
In the figure, three haloes at snapshot $z_{i+1}$ are merged into their descendant at $z_{i}$ and it enables us to calculate $z_e$ when they cross the observer's past lightcone from their positions and velocities. 
However, we cannot find a correspondence between $z_i$ and $z_{i+1}$ for a small number of haloes due to the disappearance of less-massive haloes \citep[see Section~3.5.4 of][]{behroozi13}. 
We exclude such haloes from our full-sky catalogues even though they might cross the observer's past lightcone because when it occurs is unclear. 
When a descendant halo possesses multiple progenitors, we calculate $t_{e}$ for each progenitor. 
We exclude a smaller progenitor halo that merged before crossing the lightcone, assuming that it has been merged into another more massive progenitor  (small blue circle in Figure\ref{fig:halo_crossing_time}).  
On the other hand, the one that crossed the lightcone before merging (represented by the medium red circle) is kept in our mock halo catalogue. 
The positions and velocities of haloes that cross the observer's past lightcone are interpolated by calculating the time $t_e$ at which $ds^{2}=0$.\footnote{
In the actual analysis, following \citet{hollowed19}, we use a dimensionless programme time denoted by $\tilde{t}$ instead of the cosmic time $t$. It is defined by 
\begin{equation}
\tilde{t} = a(t)^{\alpha}, 
\label{eq:pt}
\end{equation}
where we adopt $\alpha=1.0$ following \citet{hollowed19}. 
The halo velocities at the scale factor $a(t)$, $v(t)$, are accordingly written in terms of the programme time $\tilde{t}$, $v(t) = \frac{dr}{dt} = \frac{dr}{d\tilde{t}} \frac{d\tilde{t}}{dt} = v(\tilde{t}) \dot{a}\alpha a^{\alpha-1}$. 
Hence we obtain the conversion relation of the halo velocity between $\tilde{t}$ and $t$ as 
\begin{equation}
v(\tilde{t}) 
= \frac{v(t)}{\dot{a}\alpha a^{\alpha-1}} 
= \frac{v(t)}{H \alpha a^{\alpha}}, 
\label{eq:v_pt2}
\end{equation}
where $H(a)$ is the Hubble parameter defined as $H(a) = \dot{a}/a$. 
The same relation is also used for converting the speed of light. 
}

The lightcone crossing time $t_e$ by construction satisfies $t_{i+1} \leq t_{e} < t_{i}$. 
Here we newly introduce two quantities, $\tau$ and $\delta t$, defined by
\begin{equation}
\tau = t_{i} - t_{i+1},  \quad
\delta t = t_{e} - t_{i+1}. 
\end{equation}
Then the position of each halo at $t_e$, ${\bm r_{e}}$, are linearly interpolated from that at $t_{i+1}$, ${\bm r_{i+1}}$ as 
\begin{equation}
{\bm r_e} = {\bm r_{i+1}} + {\bm v_\mathrm{av}}\delta t, 
\label{eq:r_e}
\end{equation}
where ${\bm v_\mathrm{av}}$ is the average physical velocity evaluated by ${\bm v_\mathrm{av}} = ({\bm r_{i}} - {\bm r_{i+1}})/\tau$. 
Note that dark halo masses are not interpolated between snapshots and employed the values at $z_{i+1}$ snapshot since the continuity of the dark halo mass has already lost in our discrete halo catalogues due to the reassignment process of halo masses (see Section~\ref{subsubsec:hmf}). 

\subsubsection{Computing the lightcone crossing time} \label{subsubsec:lightcone_crossing_time}
We briefly describe the procedure for calculating the lightcone crossing time $t_e$ in this subsection.
See \citet{hollowed19} and \citet{korytov19} for more details. 

Since objects exist in the null geodesics of the observer's spacetime ($ds^{2} = 0$) when they cross the observer's past lightcone, equation~(\ref{eq:flrw_metric}) can be rewritten as: 
\begin{equation}
\begin{split}
|{\bm r_e}| &= \int_{t_e}^{t_0}dt\frac{c}{a(t)} \\
                  &= \int_{t_e}^{t_i}dt\frac{c}{a(t)} + \int_{t_i}^{t_0}dt\frac{c}{a(t)} \\
                  &= \int_{t_e}^{t_i}dt\frac{c}{a(t)} + |{\bm r_{i}}|, 
\end{split}
\label{eq:expand_r_e}
\end{equation}
where $c$ represents the speed of light. 
By converting the integral variable from $t$ into $t^{\prime} = t - t_{i+1}$ and expanding the scale factor to the first order, the first term of the right-hand side of equation~(\ref{eq:expand_r_e}) can be written as:
\begin{equation}
\begin{split}
\int_{t_e}^{t_i}dt\frac{c}{a(t)} &= \int_{\delta t}^{\tau}dt^{\prime}\frac{c}{a(t_{i+1} + t^{\prime})} \\
                                             &\approx \int_{\delta t}^{\tau}dt^{\prime}\frac{c}{a_{i+1} + \dot{a}_{i+1}t^{\prime}}. 
\end{split}
\label{eq:t_to_tdash}
\end{equation}
Furthermore, the right-hand side of equation~(\ref{eq:t_to_tdash}) can be analytically evaluated by approximating up to the second order in $\delta t$ and $\dot{a}$ as:
\begin{equation}
\int_{\delta t}^{\tau}dt^{\prime}\frac{c}{a_{i+1} + \dot{a}_{i+1}t^{\prime}} \approx \frac{c}{a_{i+i}} \left[ (\tau - \delta t) - \frac{1}{2}\frac{\dot{a}_{i+1}}{a_{i+1}} (\tau^{2} - \delta t^{2})\right]. 
\label{eq:2nd_order_dt_adot}
\end{equation}
We can also expand ${\bm r_{e}}$ to the second order in $\delta t$ using equation~(\ref{eq:r_e}) as:
\begin{equation}
\begin{split}
|{\bm r_e}| &= \sqrt{({\bm r_{i+1}} + {\bm v_\mathrm{av}}\delta t) \cdot ({\bm r_{i+1}} + {\bm v_\mathrm{av}}\delta t)} \\
                 &= |{\bm r_{i+1}}| \sqrt{1 + \frac{2({\bm r_{i+1}} \cdot {\bm v_\mathrm{av}})}{|{\bm r_{i+1}|}}\delta t + \frac{{\bm v_\mathrm{av} \cdot {\bm v_\mathrm{av}}}}{{\bm |{\bm r_{i+1}}|^{2}}}\delta t^{2}} \\
                 &\approx |{\bm r_{i+1}}| + \frac{{\bm r_{i+1}}\cdot{{\bm v_\mathrm{av}}}}{|{\bm r_{i+1}|}} \delta t - \frac{1}{2} \frac{({\bm r_{i+1}}\cdot{{\bm v_\mathrm{av}}})^{2}}{|{\bm r_{i+1}}|^{3}} \delta t^{2} + \frac{1}{2} \frac{{\bm v_\mathrm{av}}\cdot{\bm v_\mathrm{av}}}{|{\bm r_{i+1}}|} \delta t^{2}. 
\end{split}
\label{eq:norm_r_e}
\end{equation}
Combining equation~(\ref{eq:expand_r_e})-(\ref{eq:norm_r_e}), we obtain quadratic equation for $\delta t$:
\begin{equation}
\begin{split}
\left[ \frac{1}{2} \frac{c\dot{a}_{i+1}}{a_{i+1}^{2}} + \frac{1}{2} \left( \frac{({\bm r_{i+1}} \cdot {\bm v_\mathrm{av}})^{2}}{|{\bm r_{i+1}}|^{3}} - \frac{{\bm v_\mathrm{av}}\cdot{\bm v_\mathrm{av}}}{|{\bm r_{i+1}}|} \right) \right] \delta t^{2} \\
- \left( \frac{c}{a_{i+1}} + \frac{{\bm r_{i+1}}\cdot{{\bm v_\mathrm{av}}}}{|{\bm r_{i+1}|}} \right) \delta t \\
+ |{\bm r_{i}}| - |{\bm r_{i+1}}| + \frac{c\tau}{a_{i+1}}\left( 1 - \frac{\dot{a}_{i+1}\tau}{2a_{i+1}} \right) = 0.
\end{split}
\label{eq:delta_t}
\end{equation}
Solving equation~(\ref{eq:delta_t}) using the quadratic formula, we can evaluate the lightcone crossing time $t_{e}$ of each halo via $\delta t$. 
It is noted that haloes with the lightcone crossing time that do not satisfy the condition, $t_{i+1} \leq t_{e} < t_{i}$, are excluded from our halo catalogues at $z_{i} < z_e \leq z_{i+1}$. 

\subsubsection{Placing haloes onto the celestial sphere} \label{subsubsec:place_haloes_on_celestial_sphere}

After calculating the lightcone crossing time and interpolating positions and physical velocities of the haloes, we assign positions of the haloes on the celestial sphere and their redshift from their Cartesian coordinates. 
Right ascension (RA) and declination (Decl.) of each halo are defined as follows: 
\begin{equation}
{\rm RA} = \arctan{\left(\frac{y_{e}}{x_{e}}\right)}, 
\label{eq:ra}
\end{equation}
and
\begin{equation}
{\rm Decl.} = \arcsin{\left(\frac{z_{e}}{|{\bm r_{e}}|}\right)}, 
\label{eq:decl}
\end{equation}
where ${\bm r_{e}} = (x_{e}, y_{e}, z_{e})$ is a position of a halo within a replicated box in a Cartesian coordinate. 
We then compute redshifts of haloes, 
\begin{equation}
z = z^{\rm r} + \frac{{\bm v_{e}} \cdot {\bm r_{e}}}{|{\bm r_{e}}|} \frac{1 + z^{\rm r}}{c}, 
\label{eq:z_rsd}
\end{equation}
where ${\bm v_{e}}$ is the physical velocity of haloes at the lightcone crossing time \citep{white14} and $z^{\rm r}$ is the redshift without the effect of the peculiar velocity, evaluated from the comoving distance from the observer, $|{\bm r_{e}}|$. 

Note that in some previous studies, lightcone mock catalogues are generated by arranging periodic boxes after rotating them in order to avoid repeated structures in the final lightcone catalogues at the expense of discontinuities near the box boundaries \citep[cf.,][]{blaizot05,bernyk16}. 
However, we do not do this operation, since our simulation boxes are large enough not to find the same structure many times within the redshift range of interest, $0.3 \leq z \leq 1.4$. 

\subsection{Trimming to the HSC footprint} \label{subsec:trim_footpring}

We have created full-sky mock lightcone halo catalogues. 
We now trim some part of the full-sky catalogues to match the survey geometries of the HSC Wide layer by pixelising the celestial sphere using the {\sc HealPy}\footnote{https://github.com/healpy} package \citep{zonca19}, which is based upon the {\sc HEALPix}\footnote{https://healpix.sourceforge.io/} algorithm \citep{gorski05} with the Python framework. 
We set the resolution of the pixel to be $N_{\rm side}=4,096$, corresponding to the resolution of $2.049 \times 10^{-4}$ deg$^{2}$/pixel. 

The total area of the HSC SSP S20A Wide layer is $\sim 830$ deg$^{2}$ and the effective area covering the LRG distribution is $\sim 780$ deg$^{2}$ after applying conservative masks. 
The masks on the survey footprint are treated to avoid the effects of bright stars, bad pixels, saturated pixels, and cosmic rays. 
Our mock halo catalogues are applied with the same masks that are used in the observational study \citep{ishikawa21}. 

\subsection{HSC LRG lightcone mock catalogues} \label{subsec:mk_mockgal}
\subsubsection{Populating haloes with LRGs} \label{subsubsec:paint_lrg}

After trimming the survey footprints, we assign LRGs to the mock halo catalogues by combining the HOD and the SHAM techniques. 
The actual steps for creating mock LRG catalogues from halo catalogues are as follows. 
i) Identifying haloes that possess central and satellite LRGs from the HOD formalism. 
ii) After assigning LRGs to haloes, stellar masses are given to the LRGs by the SHAM method and their photometric redshifts are determined by considering both their redshifts and stellar masses. 
The details of each process are described below. 

First, we assign LRGs to haloes according to the HOD model shown in equations~(\ref{eq:Nc}) and (\ref{eq:Ns}). 
We use the HOD parameters constrained by \citet{ishikawa21} and Section~\ref{subsec:obs_hod_high-z} in this paper, allowing the $1\sigma$ scatter around the best-fitting model. 
Central LRGs are stochastically placed on the primary haloes according to the expectation numbers evaluated from the central occupation function, whereas satellite LRGs are assigned to subhaloes that are bound to the same primary haloes in decreasing order of maximum circular velocities $V_{\rm max}$. 
We start by populating haloes with central and satellite LRGs using the HOD model for the most massive stellar mass samples ($\log_{10}{(\Mstar/h^{-2}\Msun}) \geq 11.0$). 
We subsequently apply the HOD model for the second most massive mass samples ($\log_{10}{(\Mstar/h^{-2}\Msun}) \geq 10.75$) excluding the haloes that are already populated with LRGs. 
We repeat this procedure down to the lowest mass samples, $\log_{10}{(\Mstar/h^{-2}\Msun}) \geq 10.00$.
Note that we do not consider the off-centring effect, which represents offsets in the positions of central galaxies from the barycentres of haloes, and both central and satellite LRGs are assumed to reside at centres of their host (sub)haloes \citep[e.g.,][]{hikage13b}. 

After LRGs are populated by the above procedure, we determine their stellar masses using the SHAM technique. 
Both central and satellite LRGs are ranked according to their $V_{\rm max}$ and assigned stellar masses that are also sorted in decreasing order with $0.1$ dex scatter. 
Stellar masses are generated from the observed stellar-mass functions of LRGs that are evaluated using the stellar mass estimation of the CAMIRA algorithm \citep{oguri14} based upon the galaxy SEDs of a stellar population synthesis model of \citet{bc03}. 
The stellar mass assignment by the SHAM is done just after the halo selection of each stellar mass threshold by the HOD formalism; therefore, the stellar mass of each LRG satisfies the stellar mass thresholds of the subsamples. 

\subsubsection{Introducing photo-$z$ errors} \label{subsubsec:photo-z_error}

\begin{figure}
\includegraphics[width=\columnwidth,pagebox=cropbox,clip]{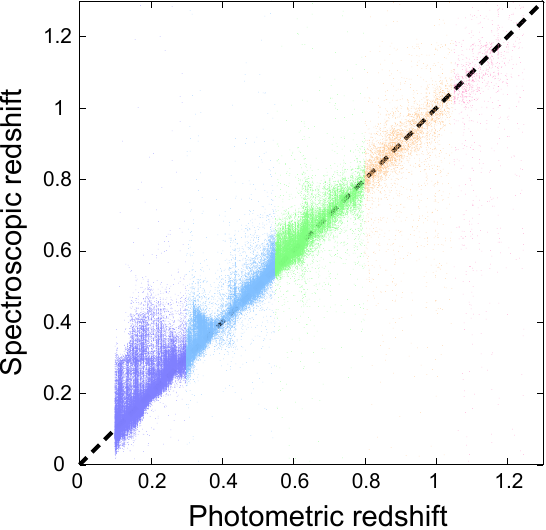}
\caption{Photometric redshifts versus spectroscopic redshifts of CAMIRA LRGs. About $5.3 \%$ of the CAMIRA LRGs have been observed by other spectroscopic observational programmes. This plot includes LRGs satisfying $\log_{10}(\Mstar/h^{-2}\Msun) \geq 10.0$. Colours indicate the difference of redshift bins of LRGs shown in Section~\ref{sec:observation}. }
\label{fig:pz-sz}
\end{figure}

We introduce photometric redshift (photo-$z$) uncertainties in our mock LRG catalogues. 
About $\sim 5.3\%$ of the entire CAMIRA LRGs are spectroscopically observed by other studies/surveys, and we can use spectroscopic redshifts of these LRGs to evaluate photo-$z$ uncertainties. 
Figure~\ref{fig:pz-sz} shows the comparison between photo-$z$'s and spectroscopic redshifts of our CAMIRA LRG samples. 

\begin{figure}
\includegraphics[width=\columnwidth]{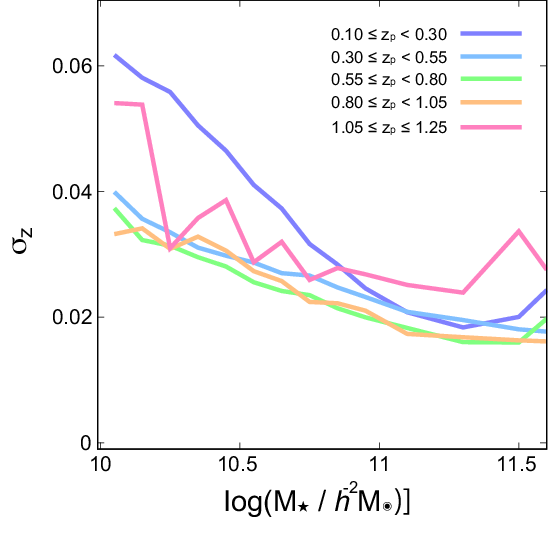}
\caption{Relation between LRG stellar masses and $1\sigma$ scatter of photometric redshifts, $\sigma_{z}$, of each redshift bin. We select LRGs that satisfy $\log_{10}(\Mstar/h^{-2}\Msun) \pm 0.05$ from each redshift bin and determine $\sigma_{z}$ by calculating root mean squares of each dispersion between photometric redshifts and spectroscopic redshifts shown in Figure~\ref{fig:pz-sz}.} 
\label{fig:sigma-z_Mstar}
\end{figure}

We assume a Gaussian distributions with the $1\sigma$ width from the measured scatter. 
We do it separately for each of the different redshift bins (equation~\ref{eq:zbin}) and stellar mass bins of $0.1$ dex. 
The result is shown in Figure~\ref{fig:sigma-z_Mstar} and we add the uncertainties to the redshifts $z$ in equation~(\ref{eq:z_rsd}). 
The actual distributions of photo-$z$ errors may not strictly adhere to Gaussian distributions, as they are contingent upon the flux errors across all observed bands. 
\citet{ansari19} made a comparison between the redshift distributions of LSST-like mock galaxies employing Gaussian photo-$z$ errors and a photo-$z$ error model that incorporated both the probability distribution functions of photo-$z$'s and the photo-$z$ quality cut. 
The findings indicated that the redshift distributions obtained from both photo-$z$ models were almost consistent in the range $0.5 \leq z \leq 1.4$. 
Hence, we posit that the photo-$z$ errors in our mock catalogues can be reasonably approximated by a Gaussian distribution. 

The observed CAMIRA LRG catalogue also contains outliers that are characterized by catastrophic photo-$z$ errors compared to spectroscopic redshifts. 
Modelling photo-$z$ uncertainties is crucial to describe the distribution of these outlier LRGs when incorporating them into mock catalogues. 
In \citet{ishikawa21}, outliers among observed CAMIRA LRGs were identified using the following criteria: 
\begin{equation}
\frac{|z_{\rm phot} - z_{\rm spec}|}{1 + z_{\rm spec}} > 0.15,
\label{eq:photoz_outlier}
\end{equation}
where $z_{\rm phot}$ and $z_{\rm spec}$ represent the observed photometric and spectroscopic redshifts, respectively. 
As reported by \citet{ishikawa21}, $\sim 2.2 \%$ of the spectroscopically-observed CAMIRA LRGs fall into the above criteria. 
To evaluate the impact of outliers on statistical quantities, we calculate the pseudo angular power spectra (see Section~\ref{subsub:pseudo_aps}) by generating a mock LRG catalogue that incorporates $2.2 \%$ outliers with catastrophic photo-$z$ uncertainties. 
To generate the mock LRG catalogue with outliers, we reassign the photometric redshifts of $2.2 \%$ of the LRGs listed in the catalogue to randomly selected redshifts drawn from a uniform distribution between $0.30 < z < 1.25$. 
We find that the amplitudes of the power spectra would be reduced by at most $\sim 3 \%$ for $z > 0.55$, and this reduction is almost scale independent. This effect leads to only $\sim 1.5 \%$ decrease in the galaxy bias. 
Therefore, we conclude that the statistical results from the mock catalogues are significantly unchanged by the lack of outliers, and we do not incorporate the outlier LRGs into our mock catalogues. 

\section{Validity Tests of Mock Catalogues} \label{sec:mock_test}

Here we perform validity tests of the constructed mock catalogues. 
In Section~\ref{subsec:comp_simulation}, we test the angular correlations of our full-sky mock halo catalogues. 
Then in Section~\ref{subsec:comp_observation} we test those of the mock HSC LRG catalogues in the footprint of the HSC SSP S20A. 

\subsection{Comparison of full-sky halo catalogues with simulations} \label{subsec:comp_simulation}

To validate our full-sky mock halo catalogues, we use full-sky gravitational lensing mock catalogues\footnote{http://cosmo.phys.hirosaki-u.ac.jp/takahasi/allsky\_raytracing/} generated by \citet{takahashi17} and measure the angular power spectra and correlation functions. 
\citet{takahashi17} constructed $108$ full-sky halo catalogues by combining simulation boxes from full $N$-body simulations with different halo mass resolutions, and thus minimum halo masses depend upon redshift. 
The mass resolutions of \citet{takahashi17} are higher than ours at $z \geq 0.476$. 
Therefore, we compare only the two-point statistics of group/cluster-scale haloes ($\Mh \geq 2 \times 10^{13} h^{-1}\Msun$) that can be properly identified by both the simulations at all the redshift ranges. 
In this subsection we adopt a redshift binning different from those in the rest of the paper to achieve a constant resolution within each redshift bin adopted in the simulations of \citet{takahashi17}. 

Furthermore, different cosmological parameters are assumed for the two sets of full-sky mock catalogues: \citet{takahashi17} adopted a flat $\Lambda$CDM cosmology from the WMAP 9 years results \citep{wmap9}, whereas we adopt one from the Planck 2015 results \citep{planck15}. 
However, we have confirmed that the impact of the difference in the cosmological parameters is negligible in terms of our statistical measures and their uncertainty level, thus the main conclusion remains unchanged. 

\subsubsection{Angular power spectrum} \label{subsubsec:aps}
\begin{figure}
\includegraphics[width=\columnwidth]{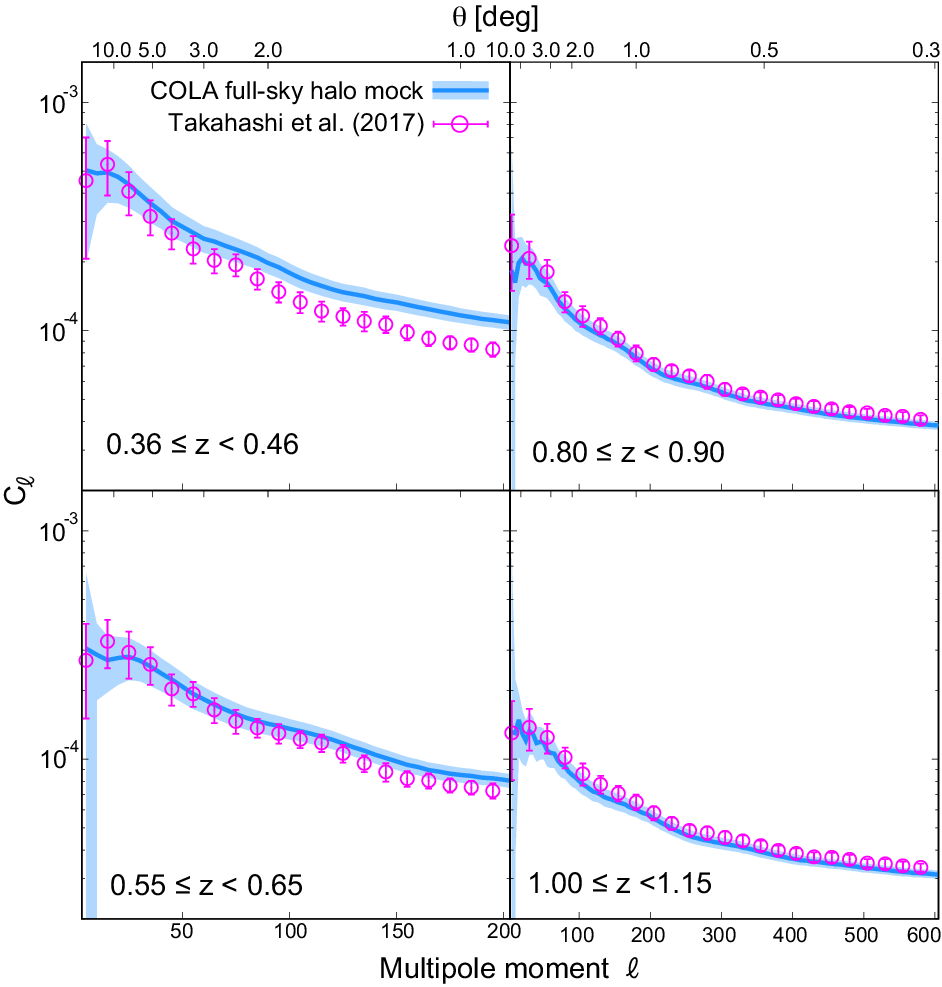}
\caption{Angular power spectra from the full-sky halo catalogues by \citet[][red circles]{takahashi17} and our COLA-based halo catalogues. The thick blue lines are the median values of angular power spectra and the shaded regions are ranges of $1\sigma$ deviations of haloes from all of $2,000$ COLA mocks. The errors of the red circles correspond to the $1\sigma$ deviations of angular power spectra of $108$ all-sky halo mocks of \citet{takahashi17}. Redshift slices are determined not to be affected by the different mass resolutions of \citet{takahashi17}. Halo mass threshold is set to $\Mh = 10^{13.5} h^{-1}\Msun$ to satisfy halo mass resolutions at all redshift shells.}
\label{fig:aps_halo}
\end{figure}

To measure the angular power spectra, 
we generate two-dimensional halo overdensity maps for each angular pixel $i$, $\delta n_i$ defined by 
\begin{equation}
\delta n_{i} = \frac{N_{i} - \bar{N}}{\bar{N}},
\label{eq:od}
\end{equation}
where $N_{i}$ is the number of haloes at the $i$-th pixel, and $\bar{N}$ is the mean number of haloes over all the pixels, $N_{\rm pix}$. 
We choose the total number of pixels to be $N_{\rm pix}=12\,\times \, 4,096^2$ to achieve the desired resolution. 
After generating the overdensity maps, we compute the angular power spectra $C_{\ell}$ of subhaloes from the full-sky $N$-body and COLA simulations, 
\begin{equation}
C_{\ell} = \frac{1}{2\ell + 1} \sum_{m=-\ell}^{\ell} |a_{\ell m}|^2,
\label{eq:c_ell}
\end{equation}
where $a_{\ell m}$ is the expansion coefficient of the spherical harmonics of the density map. 
We measure the spectra using the {\tt anafast} function in the {\sc HealPy} package. 

The resulting angular power spectra of the subhaloes are shown in Figure~\ref{fig:aps_halo}. 
We find that the angular power spectra from the COLA simulations are broadly consistent with those from the full $N$-body simulations \citep{takahashi17}. 
However, our power spectra at small angular scales at $0.3 \leq z < 0.8$ are slightly overestimated. 
The would result from the inaccuracy of the time evolution of density fields in COLA simulations. 
In the COLA simulations, the density fluctuations are guaranteed to be correct only up to the second-order in Lagrangian perturbation theory and the accuracy of the simulations will be worse when higher-order effects as well as non-perturbative effects after shell crossing are prominent, which is the case at low redshifts. 
Furthermore, the inaccuracies in the density fluctuation can accumulate towards lower redshifts. 
The matter power spectra at small angular scales from COLA simulations would be underestimated compared to those from $N$-body simulations. 
However, it is important to note that Figure~\ref{fig:aps_halo} illustrates the power spectra of biased tracers. 
In the COLA simulations, where small-scale matter clustering becomes sparser, biased tracers could exhibit enhancement, which may lead to the observed trend. 

On the other hand, the power spectra from the two full-sky mock catalogues are consistent with each other within their statistical uncertainties at large angular scales, including the BAO scales. 
These results ensure that our mocks can be used for the cosmological analysis for which the accuracy at large scales is essential. 

\subsubsection{Angular correlation function} \label{subsubsec:acf}
\begin{figure}
\includegraphics[width=\columnwidth]{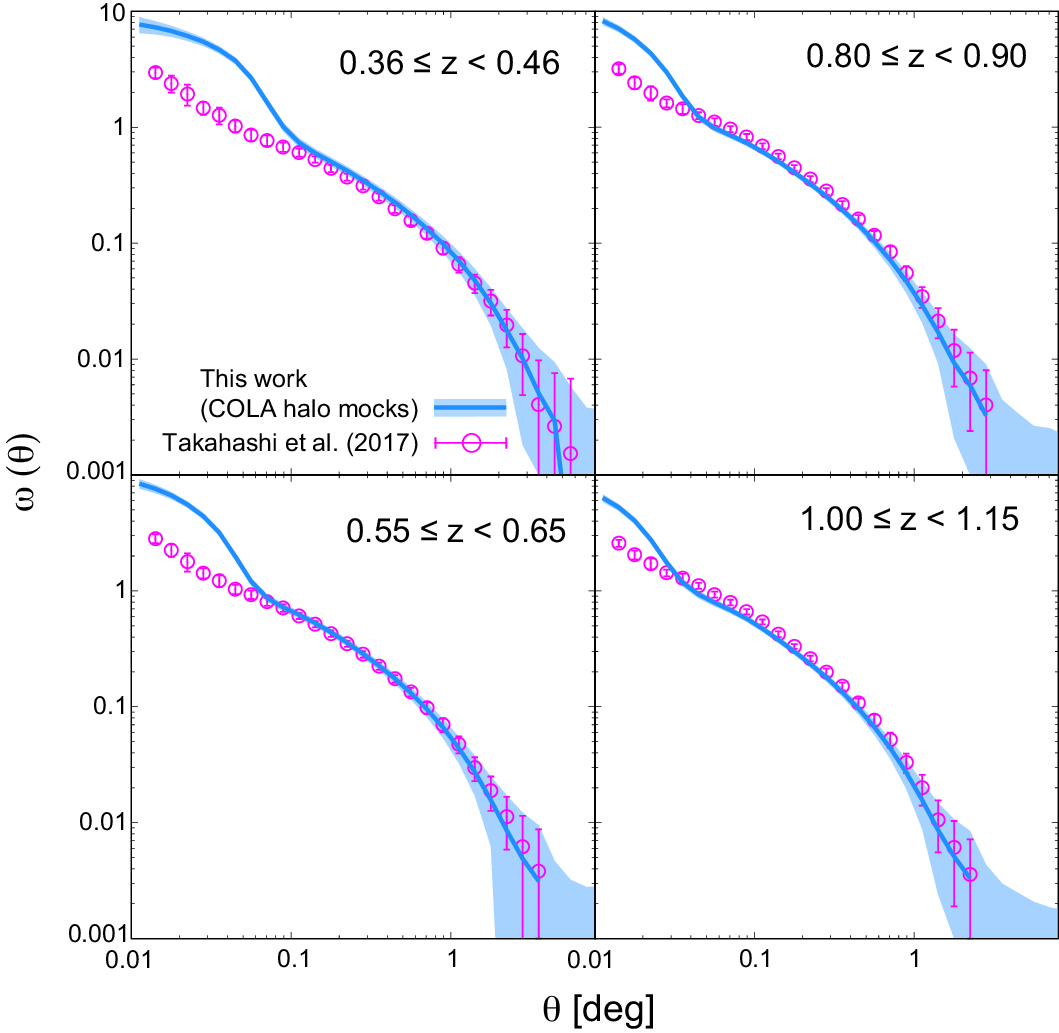}
\caption{Similar to Figure~\ref{fig:aps_halo}, but shows the comparison of angular correlation functions (ACFs). The ACFs are measured from a region that covers $2,000$ deg$^{2}$ from the mocks of the full-sky of ours and \citet{takahashi17}. }
\label{fig:acf_halo}
\end{figure}

Next, we measure the ACFs of subhaloes from the two full-sky catalogues. 
To reduce the computational time, we extract a part of the region that covers $2,000$ deg$^{2}$ from the mocks of the full-sky of ours and \citet{takahashi17}. 
We have confirmed that this is sufficient for our purpose to elucidate the difference between the two catalogues. 
The random points that cover the same regions as our catalogues are generated at a surface density $10$ times higher than that of the haloes satisfying the threshold masses. 
The ACFs are then computed using the estimator of \citet{landy93}. 

Figure~\ref{fig:acf_halo} shows the ACFs of the subhaloes measured from the $N$-body simulations of \citet{takahashi17} and COLA simulations in each redshift bin. 
There is a clear discrepancy at the 1-halo regime, since by construction the COLA algorithm is unable to correctly trace the trajectory of particles after shell-crossing. 
Overall, there are good agreements at the 2-halo regime, consistent with the angular power spectrum case,  

\subsection{Comparison of mock LRG catalogues with observation} \label{subsec:comp_observation}

In this subsection, we test the statistical properties of our HSC LRG mock catalogues against the colour-selected LRGs from the HSC SSP \citep{oguri18,ishikawa21}. 
Note that our mock LRG catalogues are generated according to the footprint of the HSC SSP S20A \citep{aihara22}, whereas the observed results are from two distinct data release; the HSC SSP S16A and S20A for the results at $0.30\leq z < 1.05$ and $1.05 \leq z \leq 1.25$, respectively (see Section~\ref{subsec:obs_hod_high-z}). 
However, they are essentially an identical population since the same selection criteria were applied to both of them \citep{oguri14,oguri18}. 

\subsubsection{Redshift distribution} \label{subsubsec:nz}
\begin{figure}
\includegraphics[width=\columnwidth]{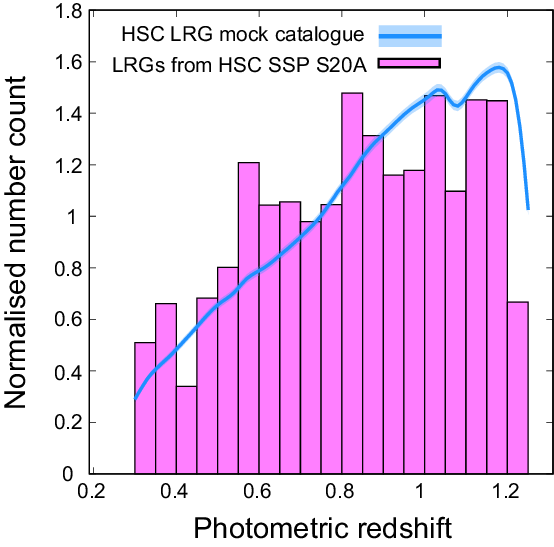}
\caption{Comparison of the redshift distribution between observed LRGs and our mock LRG catalogues. The red histogram displays the observed redshift distribution at $0.3 \leq z \leq 1.25$ derived from the CAMIRA LRG catalogue obtained from the HSC SSP S20A data, while the blue line corresponds to the median redshift distribution of our mock LRG catalogue. The blue shaded region represents the standard deviation of $2,000$ mock catalogues. }
\label{fig:nz}
\end{figure}

In Figure~\ref{fig:nz}, we compare the redshift distributions of the mock and observed LRGs. 
The red histogram is the observed redshift distribution of the LRGs obtained from HSC SSP S20A, whereas the solid blue line with shaded region represents the median and $1\sigma$ confidence interval of our $2,000$ mocks. 

Since the HOD parameters are constrained from both the observed ACFs and the number density of LRGs, the redshift distribution of our mock LRGs overall follows the observed one. 
Note that constraints on the HOD parameters were obtained for each of the five tomographic bins, and thus the redshift distribution of the mock LRGs shows discontinuity at the edges of these bins. 
However, such discontinuities are negligible as long as the statistical analysis of the LRGs is performed for sufficiently large redshift ranges. 

\subsubsection{Pseudo angular power spectrum} \label{subsub:pseudo_aps}

Unlike the full-sky analysis in Section \ref{subsubsec:aps}, we need to consider masked regions in realistic galaxy surveys. 
The angular power spectrum measured via equation~(\ref{eq:c_ell}) is biased for the incomplete sky by the mode coupling between different $\ell$'s induced by the window function, since the orthonormality of the spherical harmonic function no longer holds \citep{camacho19,alonso19}. 
A pseudo power spectrum \citep[pseudo-$C_{\ell}$;][]{hivon02} method can minimise this bias by analytically predicting the bias and correcting the effect from it \citep[see][for more details]{hivon02,hinshaw03,alonso19}. 

We use the {\sc pymaster} package\footnote{https://namaster.readthedocs.io/en/latest/}, which is a Python wrapper of the {\sc NaMaster} code \citep{alonso19}, to compute psuedo-${C}_{\ell}$ of both the observational data from the HSC SSP and our mocks. 
We choose a relatively wide band widths of $\Delta \ell = 40$ in the measurement in order to improve the S/N ratios of the pseudo-${C}_{\ell}$ for each bin. 

\begin{figure}
\includegraphics[width=\columnwidth]{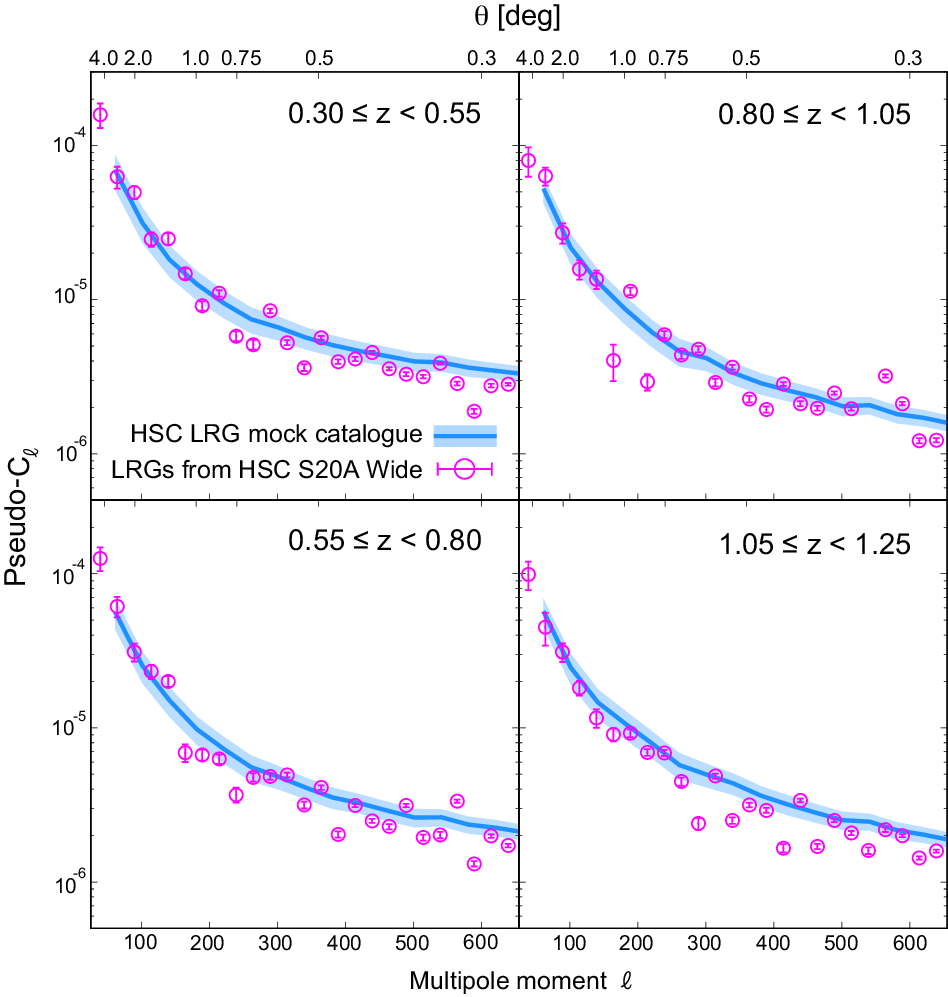}
\caption{Pseudo power spectrum (pseudo-${C}_{\ell}$) of observed LRGs and mock LRG catalogues. Red circles with errors are the pseudo-${C}_{\ell}$ of observed LRGs selected from the HSC SSP S20A Wide layer, whereas the blue solid lines with shaded area represent the median pseudo-${C}_{\ell}$ and their $1\sigma$ confidence intervals of $2,000$ HSC LRG mock lightcone catalogues. Both pseudo-${C}_{\ell}$ are calculated using the {\sc pymaster} package. }
\label{fig:pseudo_cl}
\end{figure}

Figure~\ref{fig:pseudo_cl} presents the pseudo-${C}_{\ell}$ of the observed LRGs and our HSC LRG mock catalogues. 
The errors bars in the observed pseudo-${C}_{\ell}$ are evaluated using the jackknife resampling method. 
Consistently to the full-sky halo angular power spectrum, the observed pseudo-${C}_{\ell}$'s of LRGs at large angular scale show good agreement with the predictions from the mock catalogues, demonstrating that the HOD-based LRG population successfully reconstructs the density fluctuation of the biased object over wide angular ranges. 
Interestingly, the pseudo-${C}_{\ell}$'s of the mock catalogues at small angular scales and low redshift slightly exceed those of the observed results, but the deviations are smaller compared to the full-sky halos (see Figure~\ref{fig:aps_halo}). 
This suggests that populating satellite LRGs into subhaloes can reduce the prominent small-scale clustering of the COLA mocks caused by unmerged subhaloes. 

\subsubsection{Angular corerlation function} \label{subsub:mid_acf}
\begin{figure}
\includegraphics[width=\columnwidth]{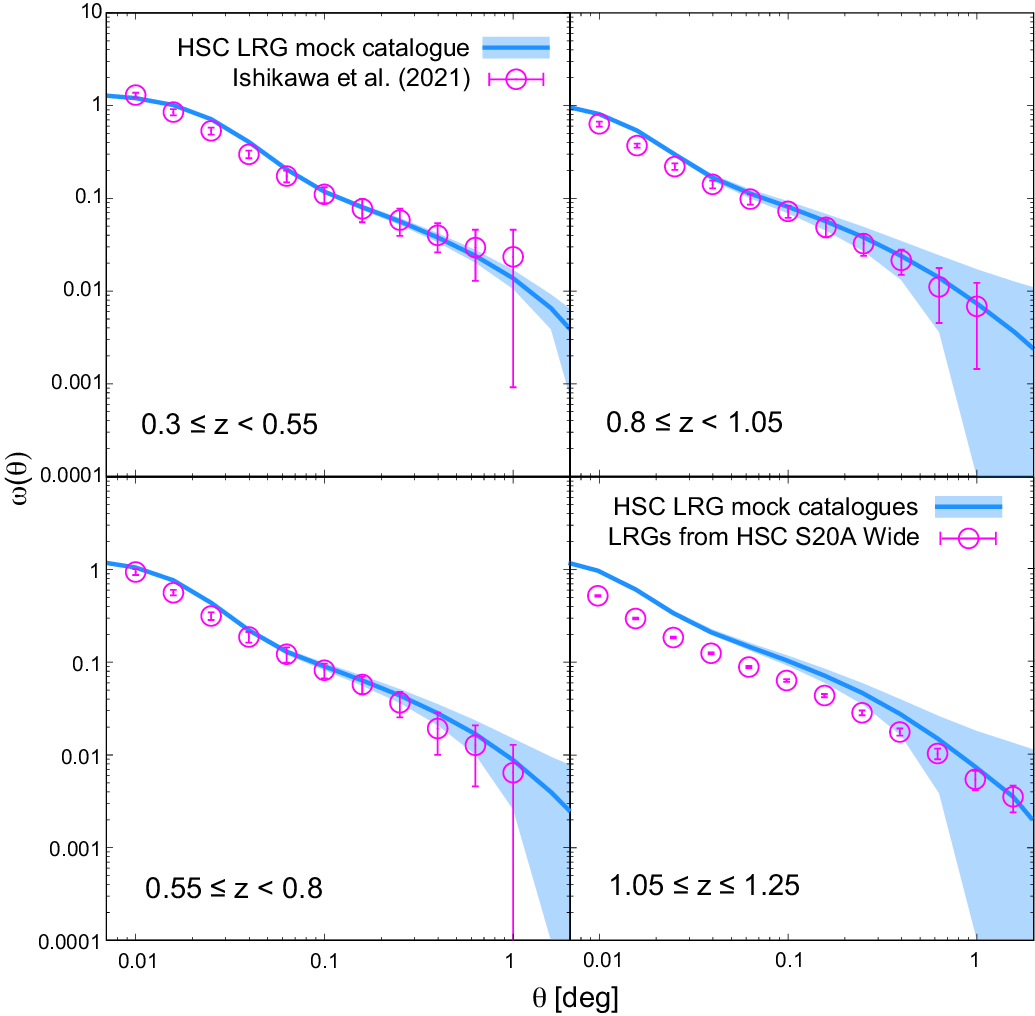}
\caption{CAMIRA LRG ACFs of the HSC SSP S16A and S20A observed by \citet[][$0.3 \leq z < 1.05$]{ishikawa21} and this study ($1.05 \leq z \leq 1.25$) and our mock HSC LRG catalogues. Thick blue lines are the median of $2,000$ total mocks and blue shaded regions are $1\sigma$ scatter derived from the $2,000$ realisations. The red circles are the observational results and the errors correspond to the $1\sigma$ scatter evaluated by the jackknife resampling method. }
\label{fig:acf_lrg}
\end{figure}

Here we measure the ACFs of galaxies, focusing on intermediate scales, i.e., $\theta \leq 1^{\circ}$. 
We study the BAO features encoded in the ACFs in the next subsection. 
Although the COLA method loses accuracy at small scales in exchange for computational speed, it is expected to give a rough test whether the mock LRGs are reasonably distributed by checking the consistency of ACFs between mocks and observation at the transition scale between the $1$- and $2$-halo terms. 

Figure~\ref{fig:acf_lrg} shows the ACFs of the observed and mock HSC LRGs for $\theta \lesssim 1^\circ$ at each redshift bin. 
The observational results at $0.3 \leq z < 1.05$ and $1.05 \leq z \leq 1.25$ are obtained using data from HSC SSP 16A in \citet{ishikawa21} and HSC SSP 20A in this work, respectively. 
We set $\Mstar = 10^{10.0} h^{-2}\Msun$ as the stellar mass threshold for all the redshift bins. 

The LRG mocks appear to reproduce the observed LRG ACFs at $0.3 \leq z < 1.05$. 
However, the ACFs of the LRG mocks are systematically in excess of observation at $1.05 \leq z \leq 1.25$. 
Several factors can explain this discrepancy. 
First, as shown in Figure~\ref{fig:sigma-z_Mstar}, the photo-$z$ scatter is considerably noisier at the $1.05 \leq z \leq 1.25$ bin when plotted as a function of the stellar mass due to the small number of spectroscopically observed LRGs in that bin. 
The transition scale from the $1$- to the $2$-halo term differs from the mass scales of galaxies, and the scatter in the photo-$z$ accuracy, which depends on the stellar masses (and also halo masses) can cause artificial noise. 
In addition, this deviation can also be attributed to the poor HOD fit in this redshift range: the ACF from the best-fitting HOD model deviates from the observed ACF at $0.05^{\circ} \lesssim \theta \lesssim 0.5^{\circ}$ and the value of the reduced $\chi^{2}$ of the best-fitting HOD exceeds $2$. 

In summary, the ACFs of our mock LRGs agree well with those observed at $z\leq 1$, but the mock LRGs fail to reproduce the spatial distributions of observed LRGs up to the intermediate scale at $z \sim 1$. 
However, the clustering at $\theta \geq 0.2^{\circ}$ is consistent with the observational results within the $1\sigma$ statistical uncertainty, meaning that our mock LRG catalogues satisfy the quality for the usage of cosmological large scale analyses. 

\subsubsection{Clustering at BAO scales} \label{subsub:bao_acf}

One of the main purposes of our LRG mocks is to evaluate covariance matrices at the BAO scales and to estimate the detectability of the BAO signature from angular clustering of the HSC LRGs. 
Here, we focus on the angular correlation function at larger scales than those in Section \ref{subsub:mid_acf}, $\theta \gtrsim 1^{\circ}$, by dividing our mocks into thin redshift shells ($\delta z \sim 0.1$) and check if the BAO bumps can be detected at the expected angular scales predicted by the fiducial cosmology \citep{planck15}. 
In the forthcoming paper, we will forecast the detectability of the BAO signals from the angular clustering of the observed HSC LRGs by mock analyses. 

\begin{figure}
\includegraphics[width=\columnwidth]{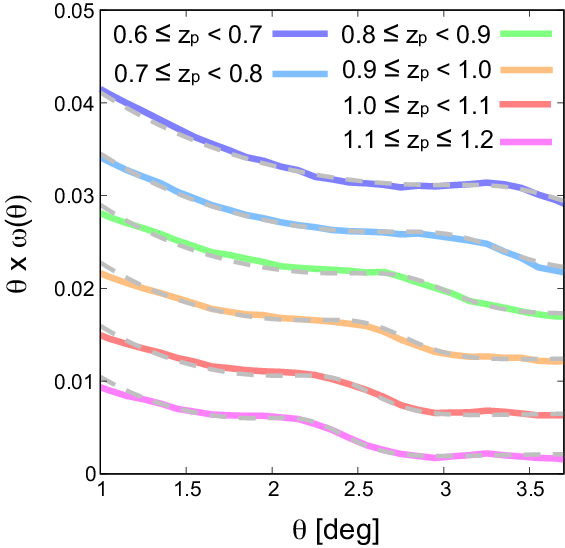}
\caption{Median ACFs at the cosmological large scale ($\theta \gtrsim 1^{\circ}$) of the total $2,000$ mock LRG catalogues by dividing into six thin redshift shells (solid lines). We select mock LRGs according to redshifts with photo-$z$ uncertainties and peculiar velocities and set $\Mstar = 10^{10.0} h^{-2}\Msun$ as a stellar-mass threshold. Dashed grey lines are obtained from the matter power spectrum by multiplying arbitrary biases at median redshifts of each shell. Amplitudes of ACFs are shifted arbitrarily for an illustrative purpose. }
\label{fig:acf_bao}
\end{figure}

Figure~\ref{fig:acf_bao} presents the ACFs of our mock LRGs at large scale for the six redshift shells. 
Each of them shows a clear BAO bump, which is in excellent agreement with the linear-theory prediction shown by the dashed grey lines. 
The solid lines for the mock LRGs show the median over a large number of mock realisations and thus the actual observation of the function from the HSC footprint would be noisier due to shot noise and cosmic variance. 

The coincidence of the BAO peaks in our LRG mocks with the predictions of the linear theory at each redshift implies the potential to detect BAO signals imprinted at the drag epoch in the ACFs even at $z>1$. 
However, the physical scales of the BAO bump are determined by multiple physical processes. 
For instance, it is well known that the BAO peak can be damped by the non-linear effects such as Silk damping \citep{silk68,seo03}. 
In addition, photometric surveys like the HSC SSP inevitably suffer from the photo-$z$ errors, which can make the BAO peaks unclear and thus difficult to determine the BAO scales at each redshift \citep[e.g.,][]{sanchez11,kishikawa23}. 
Although the uncertainties caused by photo-$z$ dominate our analyses, our mocks apparently detect BAO signals with the same scales predicted by the linear theory. 
This indicates that our mocks enable reasonable evaluation of the covariance matrices at the BAO scales, and the observed HSC LRGs can extract the cosmological parameters from the BAO-scale clustering. 

\section{Summary and Conclusion} \label{sec:summary}

We have generated a set of $2,000$ independent mock luminous red galaxy (LRG) catalogues, designed to closely resemble the observed CAMIRA LRGs obtained from the HSC SSP survey. 
These mock catalogues enable us to extract valuable cosmological information through the measurement of baryon acoustic oscillations (BAOs) at $\sim 100$ Mpc/$h$ in comoving coordinates. 

First, we have performed COmoving Lagrangian Acceleration (COLA) simulations, which are specifically designed to achieve fast computations while maintaining the accuracy of halo clustering at large scales. 
Using this method, we generated $2,000$ random realisations of matter distribution in periodic boxes. 
Dark haloes were identified from eight constant-time snapshots in the redshift range of $0.0 \leq z \leq 1.4$ using the {\sc ROCKSTAR} halo finder. 
In order to create full-sky lightcone halo catalogues, we placed the haloes that intersect the observer's past lightcone onto the celestial sphere. 
Subsequently, we trimmed these full-sky catalogues to match the footprints of the HSC SSP S20A Wide layers. 

After constructing the lightcone halo catalogues, we populated HSC-selected CAMIRA LRGs \citep{oguri18,ishikawa21} onto the simulated haloes using a combination of the halo occupation distribution (HOD) and the subhalo abundance matching (SHAM) techniques. 
The mock LRGs are assigned to haloes according to the observed HODs, and their stellar masses are determined using the SHAM technique based upon the observed CAMIRA LRG stellar-mass functions. 
We also introduced uncertainties in the photometric redshifts (photo-$z$) by incorporating the observed $\Mstar$--photo-$z$-scatter relations. 

Finally, we performed validity tests on our full-sky mock halo catalogues by comparing them with another set of full-sky halo catalogues generated by \citet{takahashi17} based on full $N$-body simulations. 
The angular power spectrum and correlation functions of our mocks at large angular scales are in good agreement with those obtained by \citet{takahashi17}. 
However, at small angular scales, the angular power spectra of our halo mocks at $z<0.8$ slightly exceed compared to those of the full $N$-body simulation, and $1$-halo terms of the angular correlation functions are more prominent. 
These differences can be attributed to the characteristics of the COLA method, where small substructures can survive after shell crossing, leading to enhanced correlations at small scales compared to the full $N$-body simulations. 

We also calculated the redshift distributions, pseudo power spectra, and angular clustering at intermediate scales and compared them with the observational results presented by \citet{ishikawa21}. 
The statistical quantities calculated from our mock catalogues showed good agreement with the observational results. 
In addition to these comparisons, we also investigated the detectability of BAO signals for the HSC SSP Wide layers using the large-scale clustering of our mock LRGs. 
Notably, the BAO features are clearly visible at scales expected from the fiducial cosmological parameters of \citet{planck15} over a wide range of redshifts. 
These results demonstrate that the HSC LRGs are capable of capturing the weak BAO signals hidden in the large-scale angular correlation functions. 
Furthermore, it is shown that our mock LRG catalogues can be utilised in large-scale cosmological analyses to calculate the covariance matrices when aiming to detect BAO signals from observed CAMIRA LRGs at various redshifts. 
In our forthcoming paper, we will discuss the detectability of BAO signals by addressing the challenges posed by photo-$z$ uncertainties through an analysis of our mock LRG catalogues. 

\section*{Acknowledgements}
We thank the anonymous referee for his/her careful reading of our manuscript and the useful comments. 
We are grateful to Masamune Oguri for providing us the CAMIRA LRG samples. 
We thank Atsushi Taruya, Satoshi Tanaka, Hironao Miyatake, and Masahiro Takada for helpful comments, discussions, and useful advice on this work. 
SI acknowledges support by JSPS KAKENHI Grant-in-Aid for Early-Career Scientists (Grant Number JP23K13145). 
SI acknowledges support of ISHIZUE 2022 of Kyoto University. 
TO acknowledges support from the Ministry of Science and Technology of Taiwan under grants Nos. MOST 111-2112-M-001-061- and NSTC 112-2112-M-001-034- and the Career Development Award, Academia Sinica (AS-CDA-108-M02) for the period of 2019-2023. 
TN was supported by JSPS KAKENHI Grant Numbers JP19H00677, JP20H05861, JP21H01081 and JP22K03634.

The Hyper Suprime-Cam (HSC) collaboration includes the astronomical communities of Japan and Taiwan, and Princeton University. The HSC instrumentation and software were developed by the National Astronomical Observatory of Japan (NAOJ), the Kavli Institute for the Physics and Mathematics of the Universe (Kavli IPMU), the University of Tokyo, the High Energy Accelerator Research Organization (KEK), the Academia Sinica Institute for Astronomy and Astrophysics in Taiwan (ASIAA), and Princeton University. Funding was contributed by the FIRST program from the Japanese Cabinet Office, the Ministry of Education, Culture, Sports, Science and Technology (MEXT), the Japan Society for the Promotion of Science (JSPS), Japan Science and Technology Agency (JST), the Toray Science Foundation, NAOJ, Kavli IPMU, KEK, ASIAA, and Princeton University. 

This paper makes use of software developed for Vera C. Rubin Observatory. We thank the Rubin Observatory for making their code available as free software at http://pipelines.lsst.io/.

This paper is based on data collected at the Subaru Telescope and retrieved from the HSC data archive system, which is operated by the Subaru Telescope and Astronomy Data Center (ADC) at NAOJ. Data analysis was in part carried out with the cooperation of Center for Computational Astrophysics (CfCA), NAOJ. We are honored and grateful for the opportunity of observing the Universe from Maunakea, which has the cultural, historical and natural significance in Hawaii. 

The Pan-STARRS1 Surveys (PS1) and the PS1 public science archive have been made possible through contributions by the Institute for Astronomy, the University of Hawaii, the Pan-STARRS Project Office, the Max Planck Society and its participating institutes, the Max Planck Institute for Astronomy, Heidelberg, and the Max Planck Institute for Extraterrestrial Physics, Garching, The Johns Hopkins University, Durham University, the University of Edinburgh, the Queen’s University Belfast, the Harvard-Smithsonian Center for Astrophysics, the Las Cumbres Observatory Global Telescope Network Incorporated, the National Central University of Taiwan, the Space Telescope Science Institute, the National Aeronautics and Space Administration under grant No. NNX08AR22G issued through the Planetary Science Division of the NASA Science Mission Directorate, the National Science Foundation grant No. AST-1238877, the University of Maryland, Eotvos Lorand University (ELTE), the Los Alamos National Laboratory, and the Gordon and Betty Moore Foundation.

Numerical computations were carried out on Cray XC50 at Center for Computational Astrophysics, National Astronomical Observatory of Japan. 
Data analysis was in part carried out on the Multi-wavelength Data Analysis System operated by the Astronomy Data Center (ADC) and the Large-scale data analysis system co-operated by the Astronomy Data Center and Subaru Telescope, National Astronomical Observatory of Japan. 
%Results of this research in part have been derived by making use of the following publicly available python packages: {\sc NumPy} \citep{harris20}, {\sc SciPy} \citep{scipy20}, {\sc mpi4py} \citep{dalcin05,dalcin11}, {\sc CosmoloPy}\footnote{http://roban.github.com/CosmoloPy/}, {\sc HealPy} \citep{gorski05,zonca19}, {\sc Numba} \citep{lam15}, {\sc Dask} \citep{rocklin15,dask16}, {\sc h5py} \citep{collette13}, {\sc NaMaster} \citep{alonso19}, and {\sc BigMPI4py} \citep{ascension20}. 

\section*{Data Availability}
The full-sky halo catalogues and the HSC LRG mock catalogues generated by this study are available upon request after a certain period of time. 

\bibliographystyle{mnras}
\bibliography{ref}

% Don't change these lines
\bsp	% typesetting comment
\label{lastpage}
%\end{comment}
\end{document}